\documentclass[%
 reprint,
showpacs,preprintnumbers,
 amsmath,amssymb,
 aps,
 prl
]{revtex4-1}

\usepackage{graphicx}
\usepackage{subfig}
\usepackage{tabularx}
\usepackage{dcolumn}
\usepackage{bm}
\usepackage{braket}
\usepackage[justification=raggedright]{caption}
\usepackage{hyperref}

\hypersetup{hidelinks}

\DeclareMathOperator{\Tr}{Tr}

\graphicspath{{./Plots/DM0/}{./DM0/}{./}}

\begin{document}

\title{On Testing Entropic Inequalities for Superconducting Qudit}

\author{Evgenii Glushkov}%
 	\email{eugene.glushkov@gmail.com}
	\affiliation{Moscow Institute of Physics and Technology}%
	\affiliation{National University of Science and Technology "MISiS"}

\author{Anastasiia Glushkova}
 \affiliation{Moscow Institute of Physics and Technology}%
 
\author{V. I. Man'ko}%
	\affiliation{Moscow Institute of Physics and Technology}%
	\affiliation{P. N. Lebedev Physical Institute, Russian Academy of Sciences}

\date{\today}

\begin{abstract}
The aim of this work is to verify the new entropic and information inequalities for non-composite systems using experimental $5 \times 5$ density matrix of the qudit state, measured by the tomographic method in a multi-level superconducting circuit. These inequalities are well-known for bipartite and tripartite systems, but have never been tested for superconducting qudits. Entropic  inequalities can also be used to evaluate the accuracy of experimental data and the value of mutual information, deduced from them, may charachterize correlations between different degrees of freedom in a noncomposite system.
\end{abstract}

\pacs{03.65.Wj, 03.67.-a, 85.25.Cp}

\maketitle

\section{\label{sec:intro}Introduction}
During the last few decades tremendous progress has been made in experimental control over quantum systems. In particular, experiments with superconducting circuits, based on Josephson junction devices \cite{dodonov1989correlated, fujii2011quantum}, have been rapidly developing recently \cite{clarke2008superconducting}. Specifically spectroscopical \cite{nakamura1997spectroscopy,averkin2014broadband} and time-domain \cite{chiorescu2003coherent}  properties of such systems were studied both theoretically and experimentally. With the improvement of coherence time of superconducting qubits it became possible to obtain the density matrices of such systems, using quantum state tomography \cite{steffen2006state} as well as Wigner tomography \cite{shalibo2013direct}.

Along with the development of quantum circuits, properties of composite quantum systems, i.e. systems containing subsystems, have been extensively studied, which resulted in numerous practical applications. These systems were also described in terms of classical information theory \cite{shannon1948note} in the quantum domain \cite{holevo2012quantum} and their information and entropic characteristics were investigated, including the von Neumann entropy and quantum mutual information, discord related measures, entropic inequalities, contextuality, causality, subadditivity and strong subadditivity conditions. 

On the contrary, the idea of using noncomposite quantum systems for quantum technologies was suggested \cite{fedorov2014entropic, kiktenko2015multilevel,kiktenko2015single} and quantum correlations in such systems have been analyzed only in recent times \cite{man2014quantum,man2014separability}. The latter opened a way of mapping information and entropic measures for composite quantum systems on the noncomposite quantum systems\cite{chernega2014tomographic, chernega2014deformed, man2014separability,man2014entanglement,man2014quantum,chernega2015no}. 

In this work, we aim to verify the entropic and information inequalities using experimental $5 \times 5$ density matrix of the qudit state ($j=2$), obtained using direct Wigner tomography in a superconducting circuit \cite{shalibo2013direct,shalibo2012control,katz2015private}.
The inequalities were obtained using approach \cite{chernega2014tomographic, chernega2014deformed, man2014separability,man2014entanglement,man2014quantum} to get analogs of subadditivity and strong subadditivity conditions, well-known for bipartite and tripartite systems, for a single qudit state. 
 
\section{\label{sec:circuits}Superconducting circuits}
Superconducting circuits with Josephson junctions are macroscopic quantum objects, that can be several micrometers wide while still preserving quantum properties. This happens because they are artficially isolated from the environment which leaves them with a single degree of freedom. The intrinsic parameters of these circuits can be engineered as desired and adjusted with an external parameter (for example, a magnetic field), so they are thereby often called "artificial atoms".

\subsection{\label{subsec:jj}Josephson junction}
The Josephson junction in superconducting circuits serves as a non-dissipative nonlinear element. It consists of two superconductors, separated by a thin insulating layer, through which Cooper-pairs can coherently tunnel. This system was described by Brian Josephson \cite{josephson1962possible}, who showed that supercurrent across the junction depends on the phase difference between the superconductors:

\vspace{-0.5cm}
\begin{equation}\label{eq:jj1}
I = I_c \sin(\phi_2-\phi_1) = I_c \sin{\phi},
\end{equation}

\noindent where $I_c$ stands for the maximum non-dissipative current flowing through the junction, i.e. the critical current. Josephson also showed that when the voltage is applied across the junction the phase difference changes in time, which leads to the oscillations of the critical current with the angular frequency $\omega$:

\vspace{-0.5cm}
\begin{equation}\label{eq:jj2}
\hbar \dot{\phi} = \hbar \omega = 2eV
\end{equation}

When we substitute this into the time derivative of Eq. \eqref{eq:jj1} and compare it to the Faraday's law, we obtain the Josephson inductance:

\vspace{-0.5cm}
\begin{equation}\label{eq:jj_inductance}
L_J (\phi)=\frac{\hbar}{2e I_c \cos{\phi}} = \frac{\Phi_0}{2\pi (I_c^2 - I^2)^{1/2}}
\end{equation}

As the Josephson junction has some intrinsic capacity $C$ it behaves as a nonlinear oscillator with angular frequency $\omega_p$:

\vspace{-0.5cm}
\begin{equation}\label{eq:plasmafreq}
\omega_p(I) = \frac{1}{\sqrt{L_J C}} = \frac{(2\pi I_c/\Phi_0 C)^{1/2}}{(1-I^2/I_c^2)^{1/4}}
\end{equation}

\vspace{-0.6cm}
\begin{figure}[ht]
\centering
\subfloat[]{
	\label{fig:fig1-a}
	\includegraphics[width=.22\textwidth]{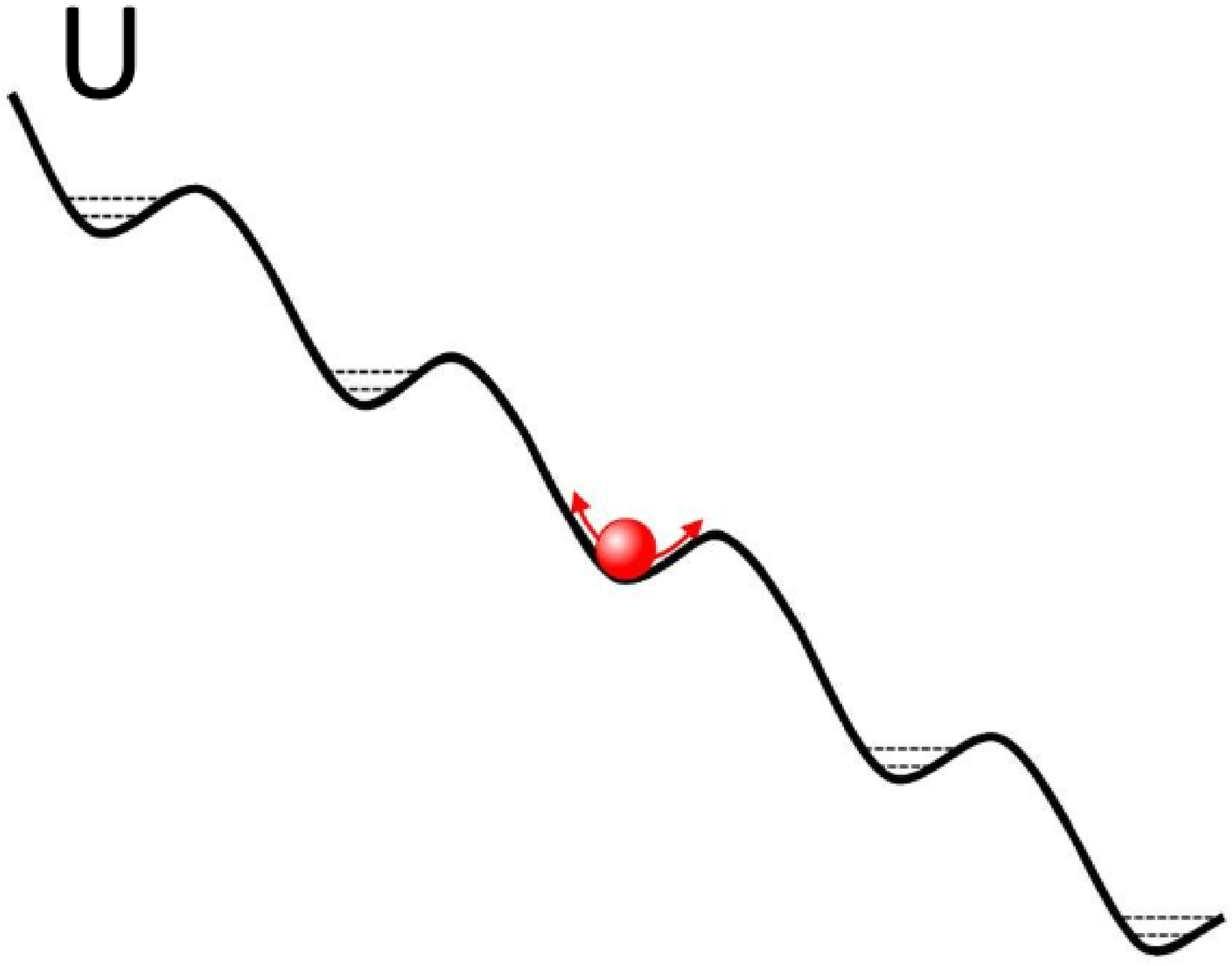}}
\subfloat[]{
	\label{fig:fig1-b}
	\includegraphics[width=.22\textwidth]{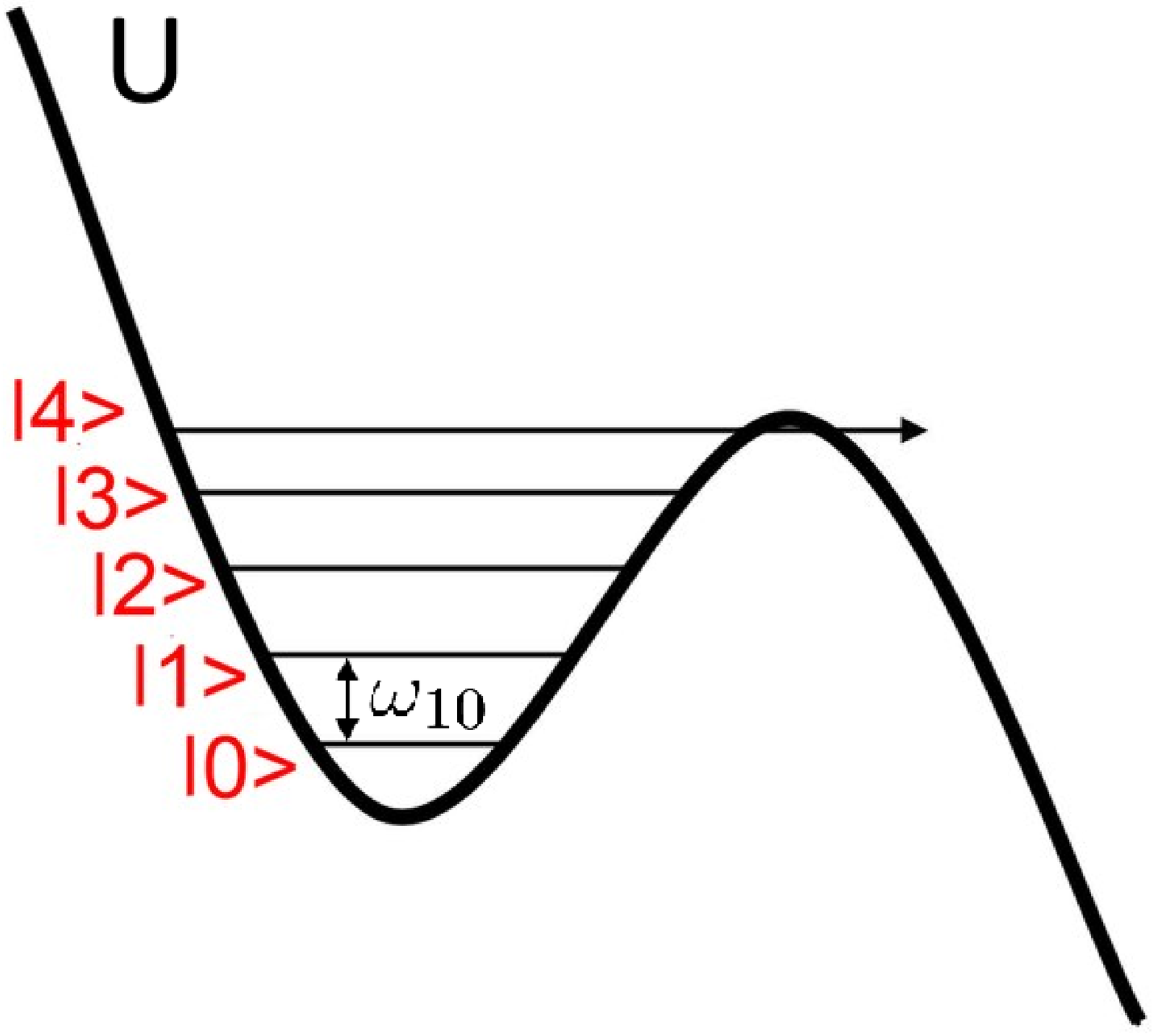}}
\caption{(color online) Tilted washboard potential (a) and quantized energy levels inside one of the potential wells (b).}
\label{fig:fig1}
\end{figure}

The total current flow trough the junction can be written as $J = I_c \sin{\phi} + V/R + C\dot{V}$. Substituting $\dot{V} = (\hbar/2e)\ddot{\phi}$ from Eq. \eqref{eq:jj2} we obtain:

\vspace{-0.5cm}
\begin{equation}\label{eq:potential1}
J = I_c \sin{\phi} +\frac{1}{R}\frac{\Phi_0}{2\pi}\dot{\phi}+C\frac{\Phi_0}{2\pi}\ddot{\phi},
\end{equation}

\noindent which is equal to the equation of motion of a particle, moving in a tilted washboard potential:

\vspace{-0.5cm}
\begin{equation}\label{eq:potential2}
m\ddot{\phi} + m \frac{1}{RC}\dot{\phi} + \frac{\partial U(\phi)}{\partial \phi} = 0,
\end{equation}

\noindent where $U = -\frac{I_c \Phi_0}{2\pi} \left(\frac{I}{I_c}\phi + \cos \phi \right)$ is shown in Fig. \ref{fig:fig1-a}.

\subsection{\label{subsec:qudit}Superconducting qudit}
\begin{figure}[ht]
\centering
\subfloat[]{
	\label{fig:fig2-a}
	\includegraphics[width=.22\textwidth]{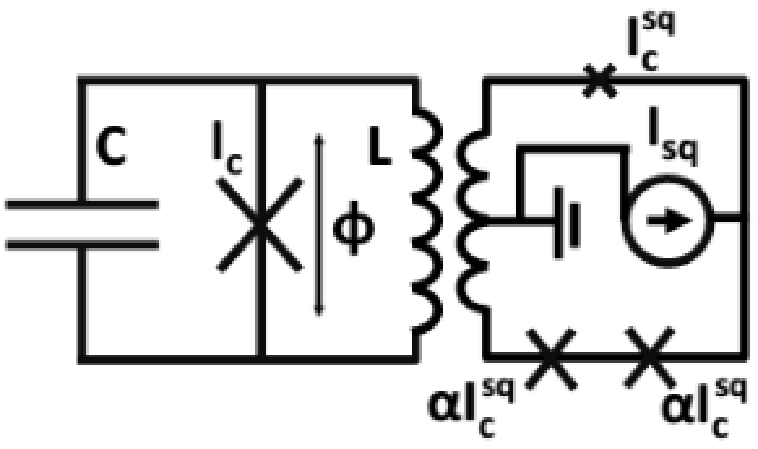}}
\subfloat[]{
	\label{fig:fig2-b}
	\includegraphics[width=.22\textwidth]{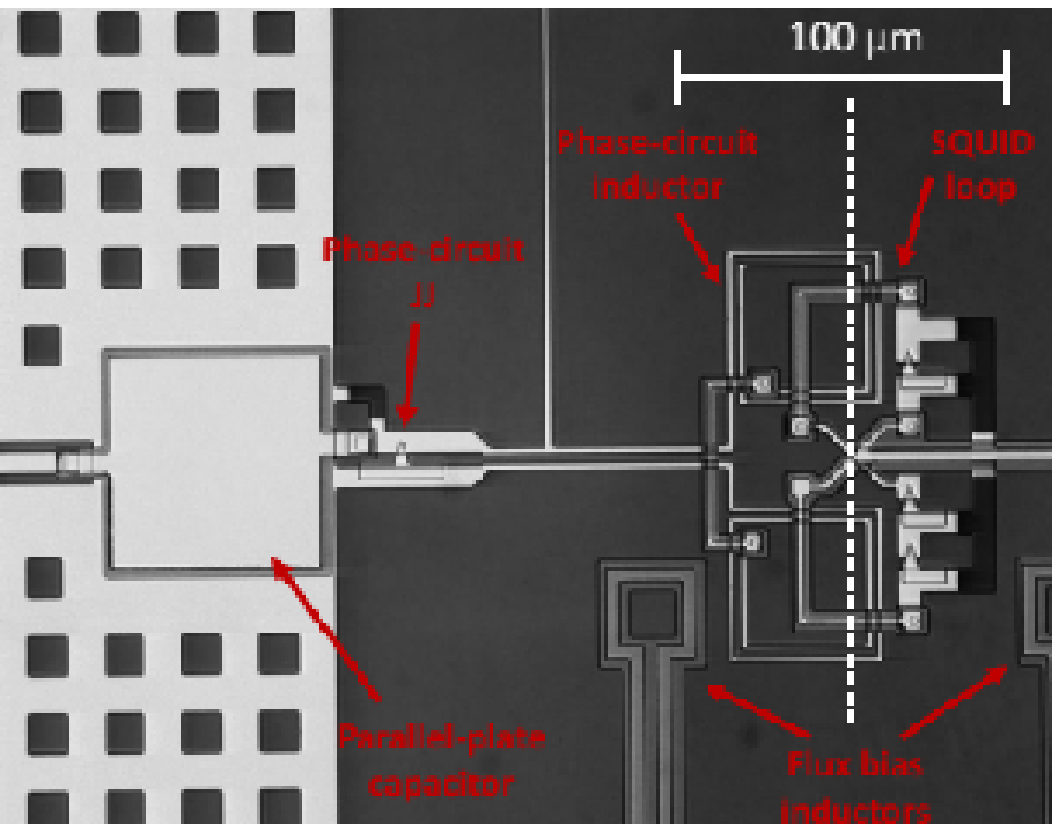}}
\caption{(color online) The Josephson phase circuit (JPC) with an on-chip SQUID.  The schematic diagram of the circuit. The left part corresponds to the JPC and the right part shows the on-chip SQUID, which is used for the readout. (b) The micrograph of the fabricated sample. \textit{(Images adapted from \cite{shalibo2012control})}}
\label{fig:fig2}
\end{figure}

A closer look at one of the wells in the tilted washboard potential in Fig.\ref{fig:fig1}(b) with the quantized energy levels gives us a perfectly suitable d-level system (qudit). Varying the potential by an external magnetic field, we can achieve a desired number of energy levels in the well. The physical implementation of this system is called the Josephson phase circuit \cite{martinis2002rabi,neeley2009emulation} and is shown in Fig. \ref{fig:fig2}.

The quantum state of the Josephson phase circuit is controlled via pulses of bias current. The measurement of the state employs the escape from the potential well via tunneling. For example, to measure the occupation probability of state $\Ket{1}$ one can pump microwaves at frequency $\omega_{41}$, which will induce a $\Ket{1} \rightarrow \Ket{4} $ transition. Then the state will rapidly tunnel due to the large tunneling rate $\Gamma_4$. When the tunneling occurs, a voltage appears across the junction, which can be measured directly by an on-chip SQUID.

In this paper we utilize the results, obtained in the experiment by Shalibo et al. \cite{shalibo2013direct, shalibo2012control, katz2015private}, in which the Wigner distribution of the Josephson phase circuit was directly measured using simple tomography pulses.

\section{\label{sec:theory}Entropic Inequalities}
Quantum states are generally described by the density matrix operator $\hat{\rho}$, which has the following properties:

\vspace{-0.5cm}
\begin{equation}
\Tr(\hat{\rho}) = 1, ~ \hat{\rho} = \hat{\rho}^{\dagger}, ~ \hat{\rho} \ge 0
\end{equation}

We consider a $5 \times 5$ density matrix for a qudit ($j = 2$):

\vspace{-0.5cm}
\begin{equation}
\rho = 
\begin{pmatrix}
\rho_{11} & \rho_{12} & \rho_{13} & \rho_{14} & \rho_{15} \\
\rho_{21} & \rho_{22} & \rho_{23} & \rho_{24} & \rho_{25} \\
\rho_{31} & \rho_{32} & \rho_{33} & \rho_{34} & \rho_{35} \\
\rho_{41} & \rho_{42} & \rho_{43} & \rho_{44} & \rho_{45} \\
\rho_{51} & \rho_{52} & \rho_{53} & \rho_{54} & \rho_{55} \\
\end{pmatrix}
\end{equation}

We can rewrite this as a $6 \times 6$ matrix, by adding one more zero row and zero column:

\vspace{-0.5cm}
\begin{equation}
\rho = \left(
\begin{array}{cccccc}
\rho_{11} & \rho_{12} & \rho_{13} & \rho_{14} & \rho_{15} & 0 \\
\rho_{21} & \rho_{22} & \rho_{23} & \rho_{24} & \rho_{25} & 0 \\
\rho_{31} & \rho_{32} & \rho_{33} & \rho_{34} & \rho_{35} & 0 \\
\rho_{41} & \rho_{42} & \rho_{43} & \rho_{44} & \rho_{45} & 0 \\
\rho_{51} & \rho_{52} & \rho_{53} & \rho_{54} & \rho_{55} & 0 \\
0 & 0 & 0 & 0 & 0 & 0 \\
\end{array}
\right)
\end{equation}

While looking at this system one can realize that it can be viewed as tensor product of two subsystems - a qubit and a qutrit. So, using an invertible mapping of indices $1 \leftrightarrow -1 ~ -1/2; 2 \leftrightarrow -1 ~ 1/2; 3 \leftrightarrow 0 ~ -1/2; 4 \leftrightarrow 0 ~ 1/2; 5 \leftrightarrow 1 ~ -1/2; 6 \leftrightarrow 1 ~ 1/2$, we obtain the density matrix, which describes the bipartite qubit-qutrit state. The density matrices of the subsystems are generally derived by taking the partial trace over the corresponding indices. We propose a simplified approach by dividing the density matrix into blocks with fewer dimensions:

\vspace{-0.5cm}
\begin{equation}\label{eq:rho-part1}
\rho = \left(
\begin{array}{ccc|ccc}
\rho_{11} & \rho_{12} & \rho_{13} & \rho_{14} & \rho_{15} & 0 \\
\rho_{21} & \rho_{22} & \rho_{23} & \rho_{24} & \rho_{25} & 0 \\
\rho_{31} & \rho_{32} & \rho_{33} & \rho_{34} & \rho_{35} & 0 \\
\hline
\rho_{41} & \rho_{42} & \rho_{43} & \rho_{44} & \rho_{45} & 0 \\
\rho_{51} & \rho_{52} & \rho_{53} & \rho_{54} & \rho_{55} & 0 \\
0 & 0 & 0 & 0 & 0 & 0 \\
\end{array}
\right)
=
\begin{pmatrix}
R_{11} & R_{12} \\
R_{21} & R_{22} \\
\end{pmatrix}
\end{equation}

Then the density matrices of the subsystems are:

\vspace{-0.5cm}
\begin{equation}\label{eq:rho1}
\rho_1 = 
\resizebox{.75\hsize}{!}{$
\begin{pmatrix}
\Tr R_{11} & \Tr R_{12} \\
\Tr R_{21} & \Tr R_{22} \\
\end{pmatrix}
= 
\begin{pmatrix}
\rho_{11}+\rho_{22}+\rho_{33} & \rho_{14}+\rho_{25} \\
\rho_{41}+\rho_{52} & \rho_{44}+\rho_{55} \\
\end{pmatrix}
$}
\end{equation}

\vspace{-0.5cm}
\begin{equation}\label{eq:rho2}
\rho_2 = 
(R_{11} + R_{22})
=
\begin{pmatrix}
 \rho_{11}+\rho_{44} & \rho_{12}+\rho_{45} & \rho_{13} \\
 \rho_{21}+\rho_{54} & \rho_{22}+\rho_{55} & \rho_{23} \\
 \rho_{31} & \rho_{32} & \rho_{33} \\
\end{pmatrix}
\end{equation}

Now we can take a look at correlations in our system. One of the most important correlation characteristics is entropy. In this work we deal with the von Neumann entropy \cite{john1955mathematical}:

\vspace{-0.5cm}
\begin{equation}
S_N = -\Tr \rho \ln{\rho}
\end{equation}

For the von Neumann entropy of the bipartite system one can write the subadditivity condition: 

\vspace{-0.5cm}
\begin{align}\label{eq:subadd1}
S_{\rho} &\le S_{\rho_1} + S_{\rho_2} \nonumber \\
-\Tr \rho \ln{\rho} &\le -\Tr \rho_1 \ln \rho_1 - \Tr \rho_2 \ln \rho_2
\end{align}

\noindent and the mutual information equals: 
\begin{equation}
I_{bp1}= S_{\rho_1} + S_{\rho_2} - S_{\rho}
\end{equation}

Now we can repeat this process for another partition of the $6 \times 6$ density matrix:

\vspace{-0.5cm}
\begin{equation}\label{eq:rho-part2}
\rho = 
\resizebox{.75\hsize}{!}{$
\left(
\begin{array}{cc|cc|cc}
\rho_{11} & \rho_{12} & \rho_{13} & \rho_{14} & \rho_{15} & 0 \\
\rho_{21} & \rho_{22} & \rho_{23} & \rho_{24} & \rho_{25} & 0 \\
\hline
\rho_{31} & \rho_{32} & \rho_{33} & \rho_{34} & \rho_{35} & 0 \\
\rho_{41} & \rho_{42} & \rho_{43} & \rho_{44} & \rho_{45} & 0 \\
\hline
\rho_{51} & \rho_{52} & \rho_{53} & \rho_{54} & \rho_{55} & 0 \\
0 & 0 & 0 & 0 & 0 & 0 \\
\end{array}
\right)
=
\begin{pmatrix}
r_{11} & r_{12} & r_{13} \\
r_{21} & r_{22} & r_{23} \\
r_{31} & r_{32} & r_{33} \\
\end{pmatrix}
$}
\end{equation}

\noindent and get the density matrices of the subsystems:

\vspace{-0.5cm}
\begin{equation} \label{eq:rhotilde1}
\tilde{\rho}_1 = 
\resizebox{.75\hsize}{!}{$
\begin{pmatrix}
\Tr r_{11} & \Tr r_{12} & \Tr r_{13} \\
\Tr r_{21} & \Tr r_{22} & \Tr r_{23} \\
\Tr r_{31} & \Tr r_{32} & \Tr r_{33} \\
\end{pmatrix}
=
\begin{pmatrix}
 \rho_{11}+\rho_{22} & \rho_{13}+\rho_{24} & \rho_{15} \\
 \rho_{31}+\rho_{42} & \rho_{33}+\rho_{44} & \rho_{35} \\
 \rho_{51} & \rho_{53} & \rho_{55} \\
\end{pmatrix}
$}
\end{equation}

\vspace{-0.5cm}
\begin{equation} \label{eq:rhotilde2}
\tilde{\rho}_2 =
\resizebox{.75\hsize}{!}{$ 
(r_{11} + r_{22} + r_{33})
=
\begin{pmatrix}
\rho_{11}+\rho_{33}+\rho_{55} & \rho_{12}+\rho_{34} \\
\rho_{21}+\rho_{43} & \rho_{22}+\rho_{44} \\
\end{pmatrix}
$}
\end{equation}

So the subadditivity condition takes the form:
\begin{align} \label{eq:subadd2}
S_{\rho} &\le S_{\tilde{\rho}_1} + S_{\tilde{\rho}_2} \nonumber \\
-\Tr \rho \ln{\rho} &\le -\Tr \tilde{\rho}_1 \ln \tilde{\rho}_1 - \Tr \tilde{\rho}_2 \ln \tilde{\rho}_2
\end{align}

\noindent and the mutual information equals: 
\begin{equation}
I_{bp2}= S_{\tilde{\rho}_1} + S_{\tilde{\rho}_2} - S_{\rho}
\end{equation}

Next we add two more zero rows and columns to this matrix to get an $8 \times 8$ matrix. The system, described by this density matrix, can be divided into three subsystems (represented by $2 \times 2$ matrices) by the following mapping of indices:
\begin{align*}
&1 \leftrightarrow -1/2 ~ -1/2 ~ -1/2; &  &2 \leftrightarrow -1/2 ~ -1/2 ~ 1/2; \\
&3 \leftrightarrow -1/2 ~ 1/2 ~ -1/2;  &  &4 \leftrightarrow -1/2 ~ 1/2 ~ 1/2;  \\
&5 \leftrightarrow 1/2 ~ -1/2 ~ -1/2;  &  &6  \leftrightarrow 1/2 ~ -1/2 ~ 1/2; \\
&7  \leftrightarrow 1/2 ~ 1/2 ~ -1/2;  &  &8  \leftrightarrow 1/2 ~ 1/2 ~ 1/2. 
\end{align*} 

Here, we use the same approach of dividing the matrix into blocks to calculate the partial traces and get the matrices for the subsystems:

\vspace{-0.4cm}
\begin{equation}
\begin{split}
\rho &= \left(
\begin{array}{cccc|cccc}
0 & 0 & 0 & 0 & 0 & 0 & 0 & 0 \\
0 & \rho_{11} & \rho_{12} & \rho_{13} & 0 & \rho_{14} & \rho_{15} & 0 \\
0 & \rho_{21} & \rho_{22} & \rho_{23} & 0 & \rho_{24} & \rho_{25} & 0 \\
0 &\rho_{31} & \rho_{32} & \rho_{33} & 0 & \rho_{34} & \rho_{35} & 0 \\
\hline
0 & 0 & 0 & 0 & 0 & 0 & 0 & 0 \\
0 & \rho_{41} & \rho_{42} & \rho_{43} & 0 & \rho_{44} & \rho_{45} & 0 \\
0 & \rho_{51} & \rho_{52} & \rho_{53} & 0 & \rho_{54} & \rho_{55} & 0 \\
0 & 0 & 0 & 0 & 0 & 0 & 0 & 0 \\
\end{array}
\right) \\
&= \left(
\begin{array}{cc|cc|cc|cc}
0 & 0 & 0 & 0 & 0 & 0 & 0 & 0 \\
0 & \rho_{11} & \rho_{12} & \rho_{13} & 0 & \rho_{14} & \rho_{15} & 0 \\
\hline
0 & \rho_{21} & \rho_{22} & \rho_{23} & 0 & \rho_{24} & \rho_{25} & 0 \\
0 &\rho_{31} & \rho_{32} & \rho_{33} & 0 & \rho_{34} & \rho_{35} & 0 \\
\hline
0 & 0 & 0 & 0 & 0 & 0 & 0 & 0 \\
0 & \rho_{41} & \rho_{42} & \rho_{43} & 0 & \rho_{44} & \rho_{45} & 0 \\
\hline
0 & \rho_{51} & \rho_{52} & \rho_{53} & 0 & \rho_{54} & \rho_{55} & 0 \\
0 & 0 & 0 & 0 & 0 & 0 & 0 & 0 \\
\end{array}
\right)
\end{split}
\end{equation}

The density matrices that we are using hereinafter are the matrix of the second subsystem, $R_2$, and two joint matrices of the "qubit-qubit" subsystems, $\rho_{12}$ and $\rho_{23}$:

 \vspace{-0.4cm}
 \begin{equation}\label{eq:rho12}
 \rho_{12}  = 
 \resizebox{0.35\hsize}{!}{$
 \begin{pmatrix}
 \rho_{11} & \rho_{13} & \rho_{14} & 0 \\
 \rho_{31} & \rho_{22}+\rho_{33} & \rho_{34} & \rho_{25} \\
 \rho_{41} & \rho_{43} & \rho_{44} & 0 \\ 
 0 & \rho_{52} & 0 & \rho_{55} \\ 
 \end{pmatrix}
 $},
  \end{equation}
  
  \vspace{-0.4cm}
  \begin{equation}\label{eq:rho23}
  \rho_{23} = 
  \resizebox{0.75\hsize}{!}{$
  \begin{pmatrix}
  0 & 0 & 0 & 0 \\
  0 & \rho_{11}+\rho_{14}+\rho_{41}+\rho_{44} & \rho_{12}+\rho_{15}+\rho_{42}+\rho_{45} & \rho_{13}+\rho_{43} \\
  0 & \rho_{21}+\rho_{24}+\rho_{51}+\rho_{54} & \rho_{22}+\rho_{25}+\rho_{52}+\rho_{55} & \rho_{23}+\rho_{53} \\
  0 & \rho_{31}+\rho_{34} & \rho_{32}+\rho_{35} & \rho_{33} \\ 
  \end{pmatrix} 
 $},
  \end{equation}
  
  \vspace{-0.4cm}
  \begin{equation}\label{eq:R2}
   R_2  = 
   \resizebox{0.75\hsize}{!}{$
   \begin{pmatrix}
   \rho_{11}+\rho_{14}+\rho_{41}+\rho_{44} & \rho_{13}+\rho_{43} \\
   \rho_{31}+\rho_{34} &  \rho_{22}+\rho_{33}+\rho_{25}+\rho_{52}+\rho_{55} \\
   \end{pmatrix}
   $}.
   \end{equation}

For this kind of tripartite system one can write the strong subadditivity condition \cite{lieb2002proof}:

\vspace{-0.3cm}
\[
S_{\rho} + S_{R_2} \le S_{\rho_{12}} + S_{\rho_{23}}
\]

\vspace{-0.8cm}
\begin{equation}\label{eq:strongsubadd}
\resizebox{0.85\hsize}{!}{$
-\Tr \rho \ln{\rho} -\Tr R_2 \ln R_2 \le  - \Tr \rho_{12} \ln \rho_{12} - \Tr \rho_{23} \ln \rho_{23}
$}
\end{equation}

\section{\label{sec:exper}Verifying experimental data}

Next, we calculate the density matrices  of the subsystems from the experimentally obtained $5 \times 5$ density matrix. This density matrix corresponds to the qudit, mentioned in section \hyperref[subsec:qudit]{Superconducting qudit}, and was measured in \cite{shalibo2013direct,shalibo2012control,katz2015private}.

\vspace{-0.15cm}
\begin{figure}[h]
\centering
\includegraphics[width=.36\textwidth]{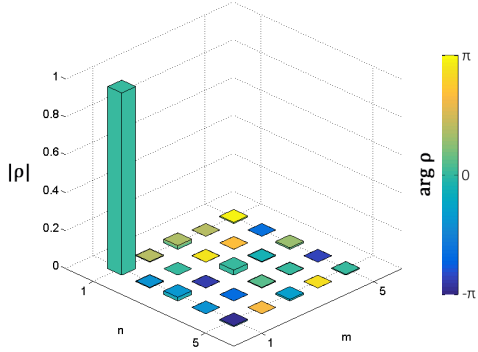}
\caption{(color online) Experimentally obtained density matrix of a superconducting qudit \cite{shalibo2013direct, shalibo2012control, katz2015private}.}
\end{figure}

The density matrices in Eq.\eqref{eq:rho1} and \eqref{eq:rho2} are:
\vspace{-0.2cm}
\begin{equation}
\rho_1 = 
\resizebox{.75\hsize}{!}{$
\left(\begin{array}{cc} 0.985 & 8.3\cdot 10^{-5} - 2.7\cdot 10^{-4}\, \mathrm{i}\\ 8.3\cdot 10^{-5} + 2.7\cdot 10^{-4}\, \mathrm{i} & 0.006 \end{array}\right)
\nonumber
$},
\end{equation}

\vspace{-0.4cm}
\begin{equation}
\rho_2 =
\resizebox{.75\hsize}{!}{$ 
\left(\begin{array}{ccc} 0.96 & 8.8\cdot 10^{-4} - 0.003\, \mathrm{i} & 0.008 - 0.018\, \mathrm{i}\\ 8.8\cdot 10^{-4} + 0.003\, \mathrm{i} & 0.004 & -7.6\cdot 10^{-4} - 2.9\cdot 10^{-4}\, \mathrm{i}\\ 0.008 + 0.018\, \mathrm{i} & -7.6\cdot 10^{-4} + 2.9\cdot 10^{-4}\, \mathrm{i} & 0.026 \end{array}\right)
\nonumber
$}.
\end{equation}

Analogously, for the Eq.\eqref{eq:rhotilde1} and \eqref{eq:rhotilde2} we obtain the following density matrices:

\vspace{-0.5cm}
\begin{equation}
\tilde{\rho}_1 =
\resizebox{.75\hsize}{!}{$
\left(\begin{array}{ccc} 0.96 & 0.008 - 0.018\, \mathrm{i} & -0.006 - 8.6\cdot 10^{-4}\, \mathrm{i}\\ 0.008 + 0.018\, \mathrm{i} & 0.028 & 0.005 - 0.007\, \mathrm{i}\\ -0.006 + 8.6\cdot 10^{-4}\, \mathrm{i} & 0.005 + 0.007\, \mathrm{i} & 0.004 \end{array}\right)
\nonumber
$},
\end{equation}

\vspace{-0.4cm}
\begin{equation}
\tilde{\rho}_2 =
\resizebox{.5\hsize}{!}{$ 
\left(\begin{array}{cc} 0.99 & 0.005 - 0.002\, \mathrm{i}\\ 0.005 + 0.002\, \mathrm{i} & 0.002 \end{array}\right)
\nonumber
$}.
\end{equation}

\vspace{-0.98cm}
\begin{figure}[h!]%
\centering
\subfloat[$\rho_1$]{%
\label{fig:msub-a}%
\includegraphics[width=.35\linewidth]{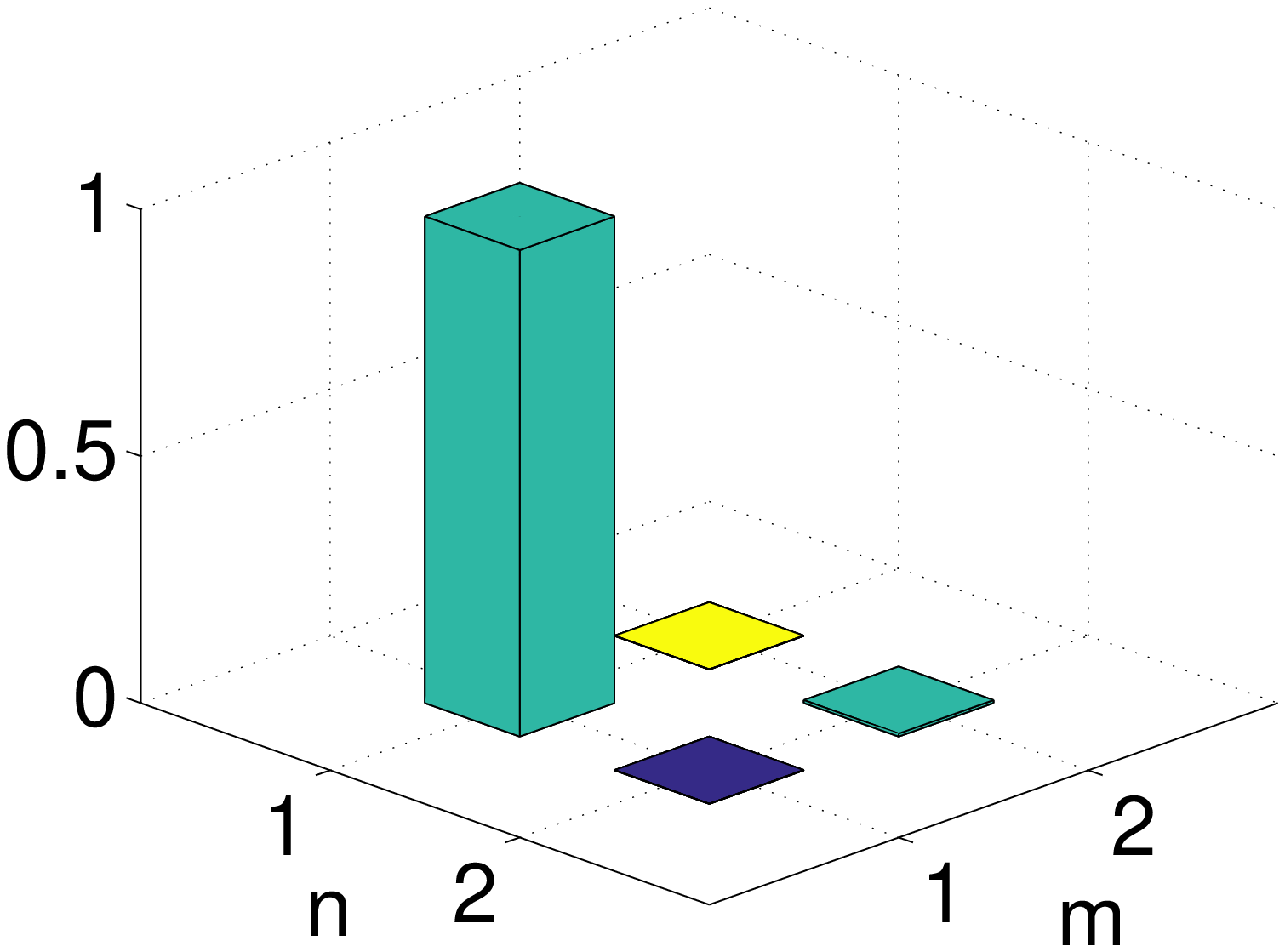}}%
\vspace{-5pt}%
\subfloat[$\rho_2$]{%
\label{fig:msub-b}%
\includegraphics[width=.35\linewidth]{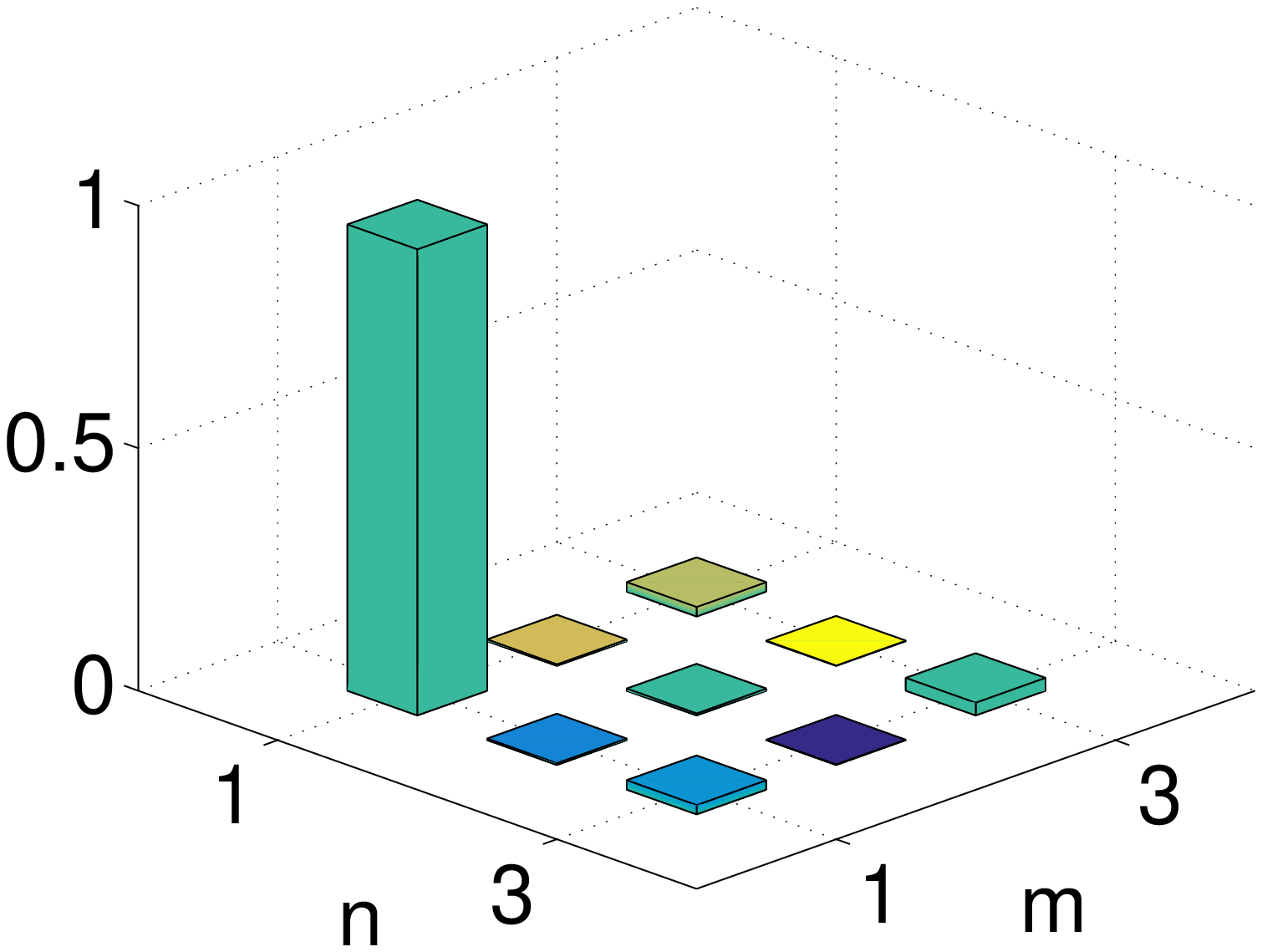}}\\
\subfloat[$\tilde{\rho}_1$]{%
\label{fig:msub-c}%
\includegraphics[width=.35\linewidth]{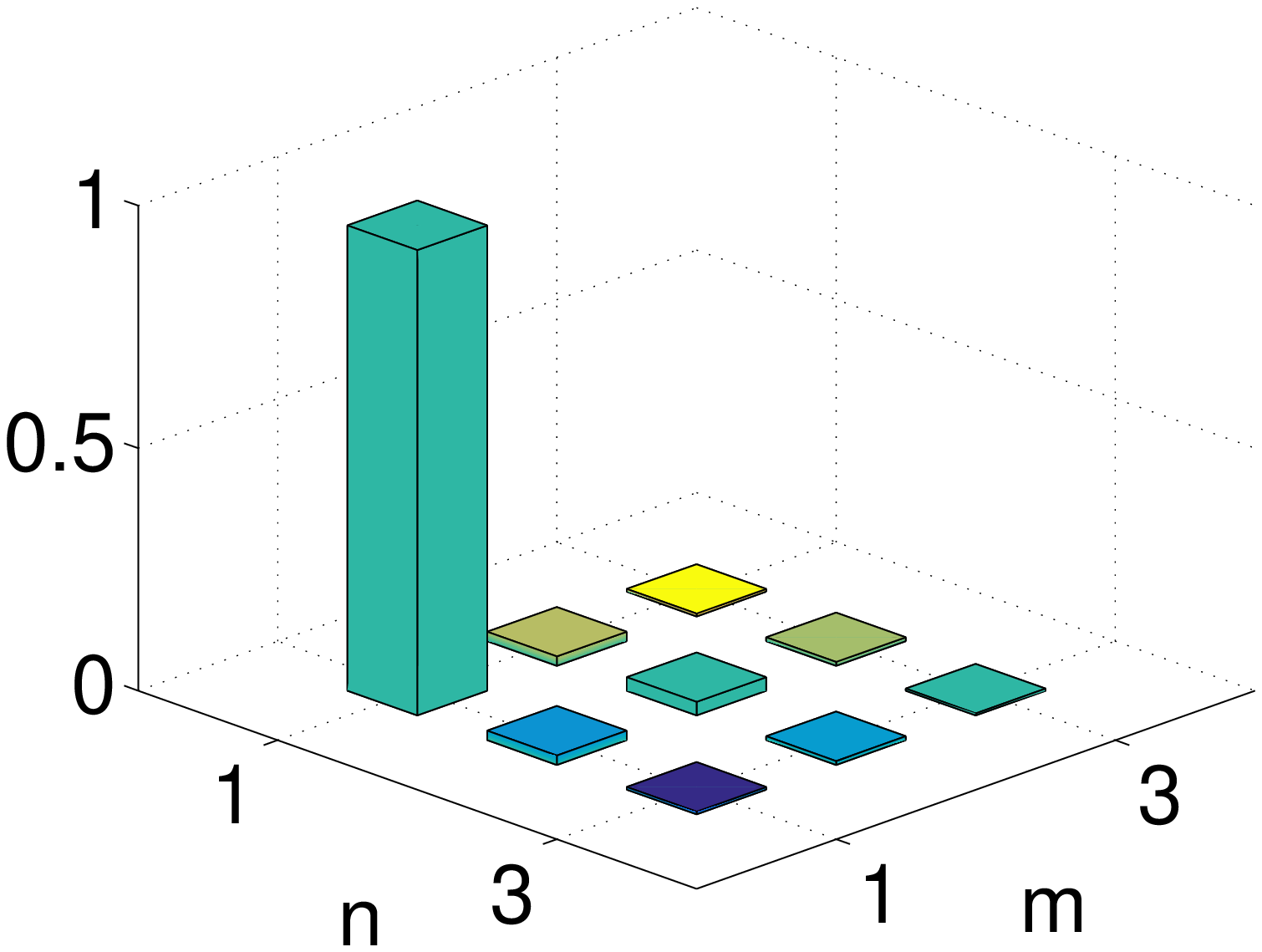}}%
\vspace{-5pt}%
\subfloat[$\tilde{\rho}_2$]{%
\label{fig:msub-d}%
\includegraphics[width=.35\linewidth]{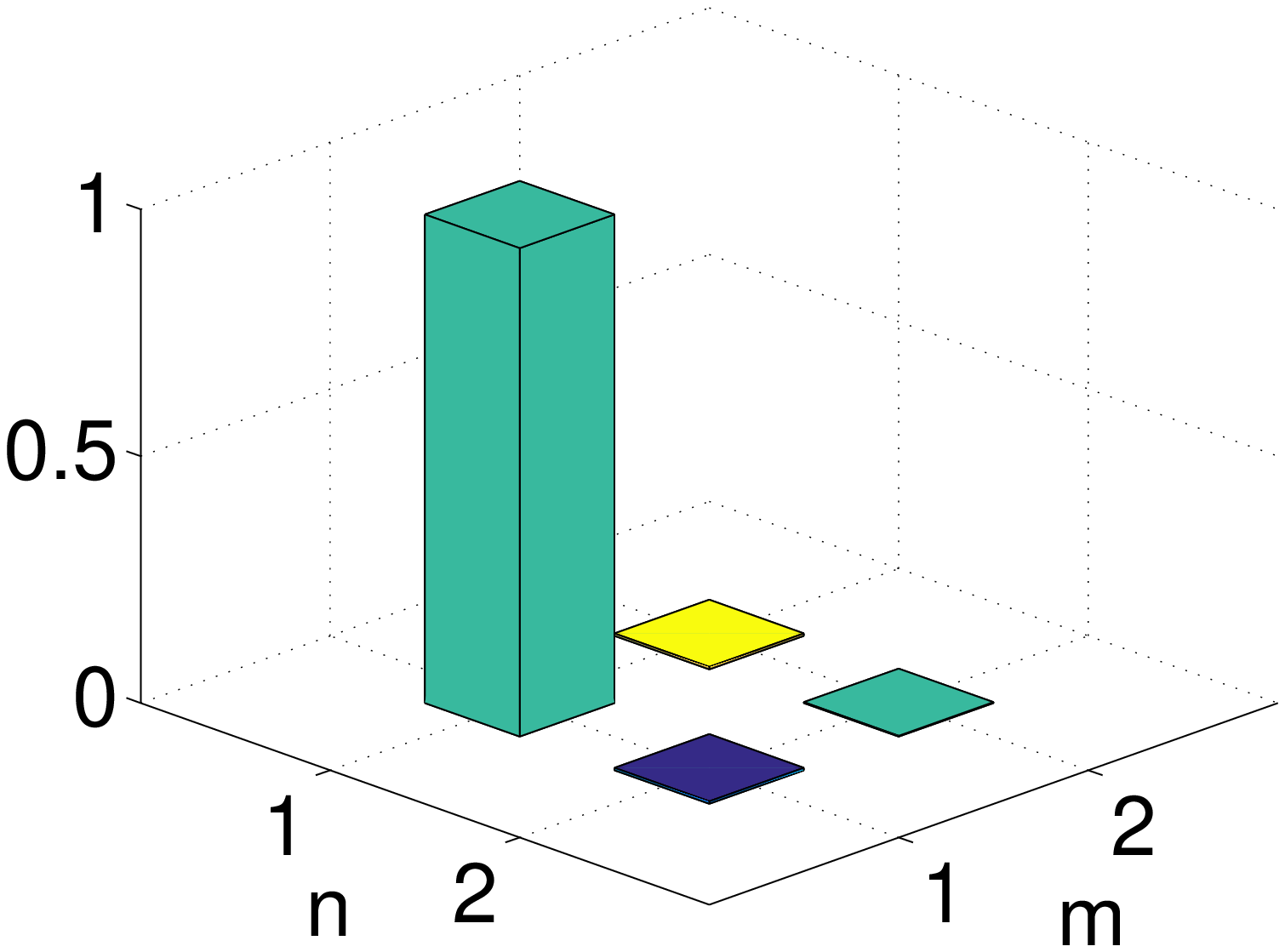}}%
\caption[A set of four sub-floats.]{(color online) Calculated density matrices for the bipartite system:
\subref{fig:msub-a}, \subref{fig:msub-b} correspond to the first partition;
\subref{fig:msub-c}, \subref{fig:msub-d} correspond to another partition.}%
\label{fig:msub}%
\end{figure}

\begin{figure}[ht]
\centering
\subfloat[First partition]{
	\label{fig:subadd-a}
	\includegraphics[width=.24\textwidth]{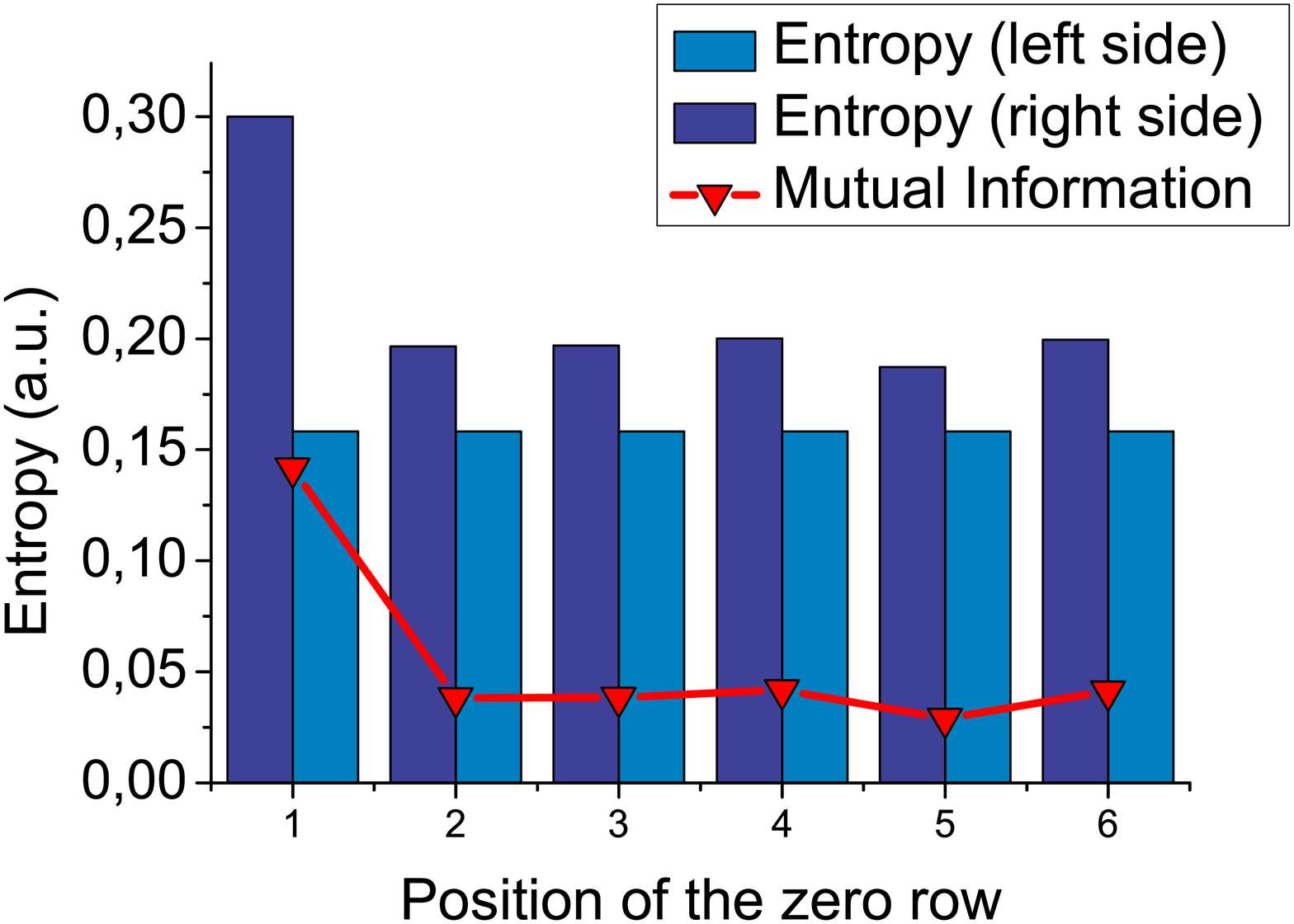}}
\subfloat[Another partition]{
	\label{fig:subadd-b}
	\includegraphics[width=.24\textwidth]{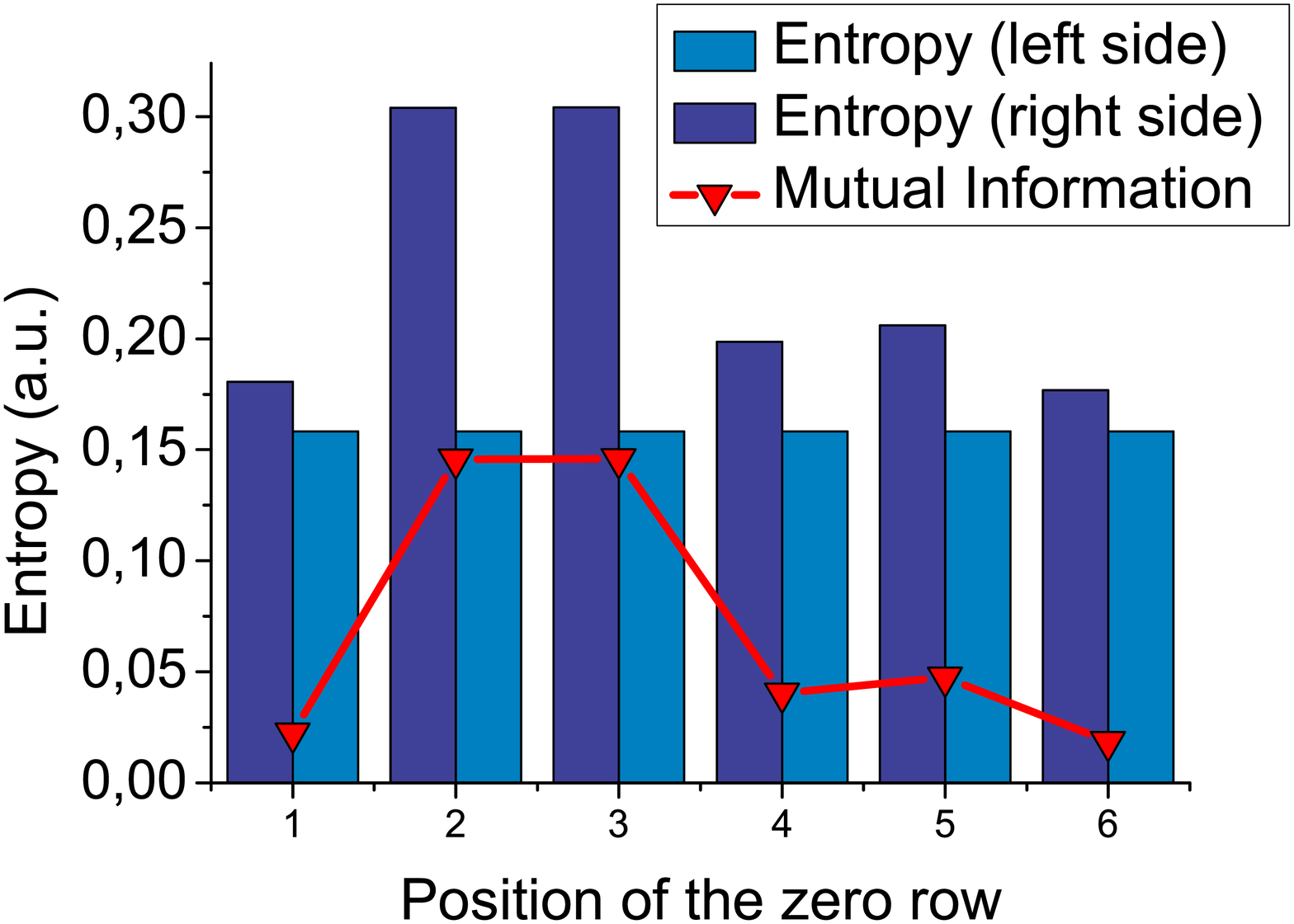}}
\caption{(color online) The entropies and mutual information from Table \ref{tabl:subadd}, plotted versus the position of the zero-row, for the "qubit-qutrit" partition.}
\label{fig:subadd}
\end{figure}

Using these matrices, we can calculate corresponding entropies and mutual information and test the subadditivity condition for different partitions in Eq.\eqref{eq:subadd1} and \eqref{eq:subadd2}. Moreover, we can also change the position of the zero-row and zero-column in Eq.\eqref{eq:rho-part1} and \eqref{eq:rho-part2} to see, how these entities will change. The results of these calculations are given in the Table \ref{tabl:subadd} and shown in the Fig. \ref{fig:subadd}.

\begin{table}[h!]
\centering
\caption{Calculated entropies and mutual information.}
\vspace{-0.3cm} 
\label{tabl:subadd}
\begin{tabular}{ c c c c c c  }
 \hline
 \hline
 Zero-row position & $S_{\rho}$ & $S_{bp1}$ & $S_{bp2}$ & $I_{bp1}$ & $I_{bp2}$ \\
 \hline
 $(1;1)$ & $0.1583$ & $0.300$ & $0.180$ & $0.1418$ & $0.0224$ \\
  \hline
  $(2;2)$ & $0.1583$ & $0.1965$ & $0.3040$ & $0.0383$ & $0.1457$ \\
 \hline
 $(3;3)$ & $0.1583$ & $0.1968$ & $0.3042$ & $0.0386$ & $0.1459$ \\
 \hline
 $(4;4)$ & $0.1583$ & $0.2001$ & $0.1987$ & $0.0418$ & $0.0404$ \\
 \hline
 $(5;5)$ & $0.1583$ & $0.1873$ & $0.2059$ & $0.0291$ & $0.0477$ \\
 \hline
 $(6;6)$ & $0.1583$ & $0.1996$ & $0.1768$ & $0.0413$ & $0.0185$ \\
 \hline
 \hline
\end{tabular}
\end{table}

\vspace{-0.7cm}

Finally, we calculate the density matrices for the tripartite system (Eq.\eqref{eq:rho12}, \eqref{eq:rho23}, \eqref{eq:R2}):
 
 \vspace{-0.3cm}
 \begin{equation}
 \rho_{12}  = 
 \resizebox{0.84\hsize}{!}{$
 \left(\begin{array}{cccc} 
 0.959 & 0.008 - 0.018\, \mathrm{i} & 0.0002 - 0.0004\, \mathrm{i} & 0\\
  0.008 + 0.018\, \mathrm{i} & 0.026 & 0.003 + 0.0013\, \mathrm{i} & -0.0001 + 0.0002\, \mathrm{i}\\ 
  0.0002 + 0.0004\, \mathrm{i} & 0.003 - 0.0013\, \mathrm{i} & 0.0018 & 0\\
  0 & -0.0001 - 0.0002\, \mathrm{i} & 0 & 0.004
 \end{array}\right)
 \nonumber
 $},
  \end{equation}
  
  \vspace{-0.3cm}
  \begin{equation}
  \rho_{23} = 
  \resizebox{0.75\hsize}{!}{$
  \left(\begin{array}{cccc} 
  0 & 0 & 0 & 0\\ 
  0 & 0.961\, \mathrm{i} & -0.005 - 0.004\, \mathrm{i} & 0.012 - 0.019\, \mathrm{i}\\ 
  0 & -0.005 + 0.004\, \mathrm{i} & 0.004 & 0.004 + 0.0064\, \mathrm{i}\\ 
  0 & 0.012 + 0.019\, \mathrm{i} & 0.004 - 0.0064\, \mathrm{i} & 0.026 
  \end{array}\right) 
  \nonumber
  $},
  \end{equation}
  
  \vspace{-0.1cm}
  \begin{equation}
   R_2  = 
   \resizebox{.5\hsize}{!}{$ 
   \left(\begin{array}{cc} 
   0.961 & 0.012 - 0.019\, \mathrm{i}\\ 
   0.012 + 0.019\, \mathrm{i} & 0.030  
   \end{array}\right)
   \nonumber
   $}.
   \end{equation}
  
\vspace{-0.3cm}
\begin{figure}[ht]
\centering
\subfloat[$R_2$]{
	\includegraphics[width=.15\textwidth]{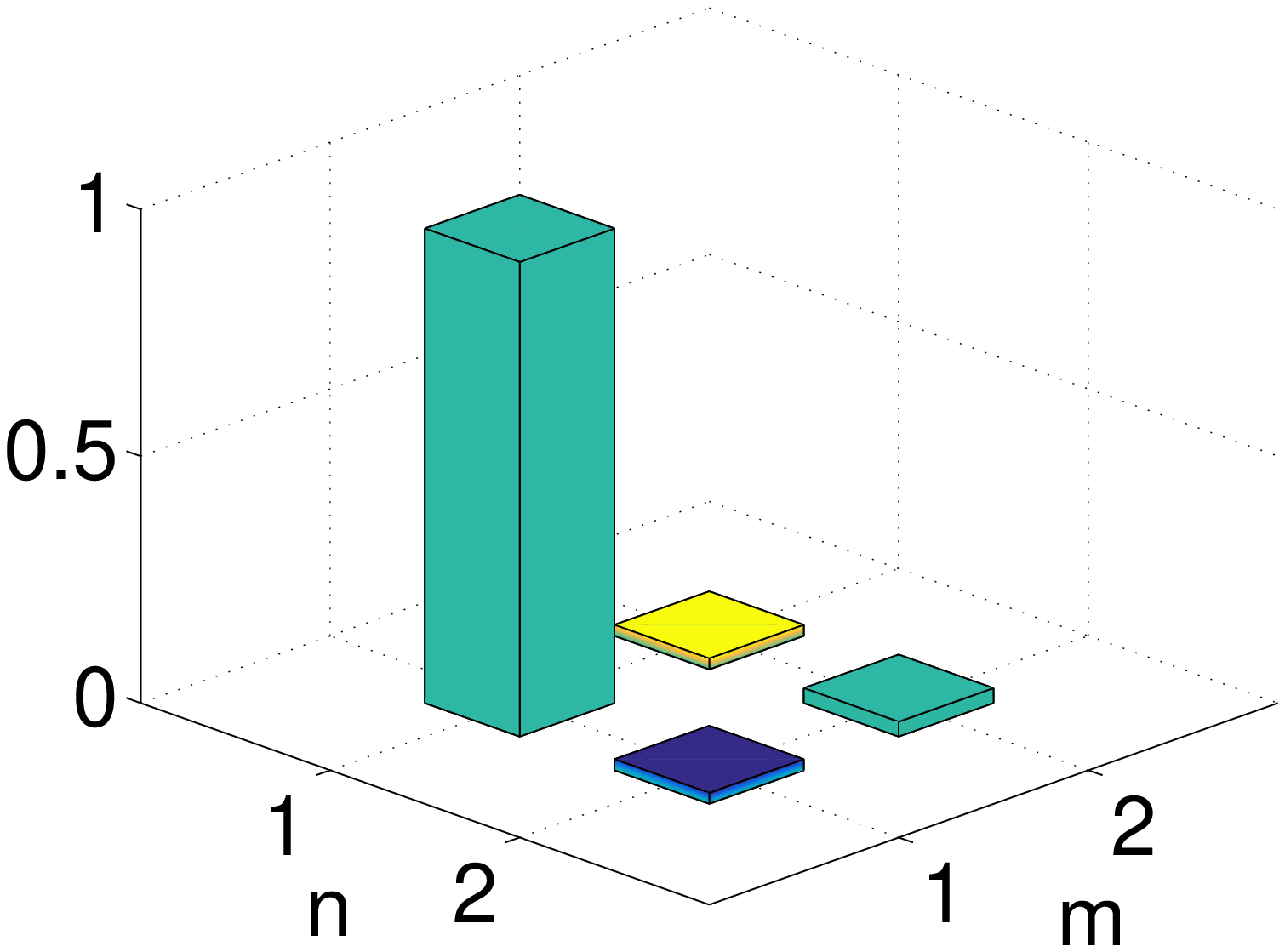}}
\subfloat[$\rho_{12}$]{
	\includegraphics[width=.15\textwidth]{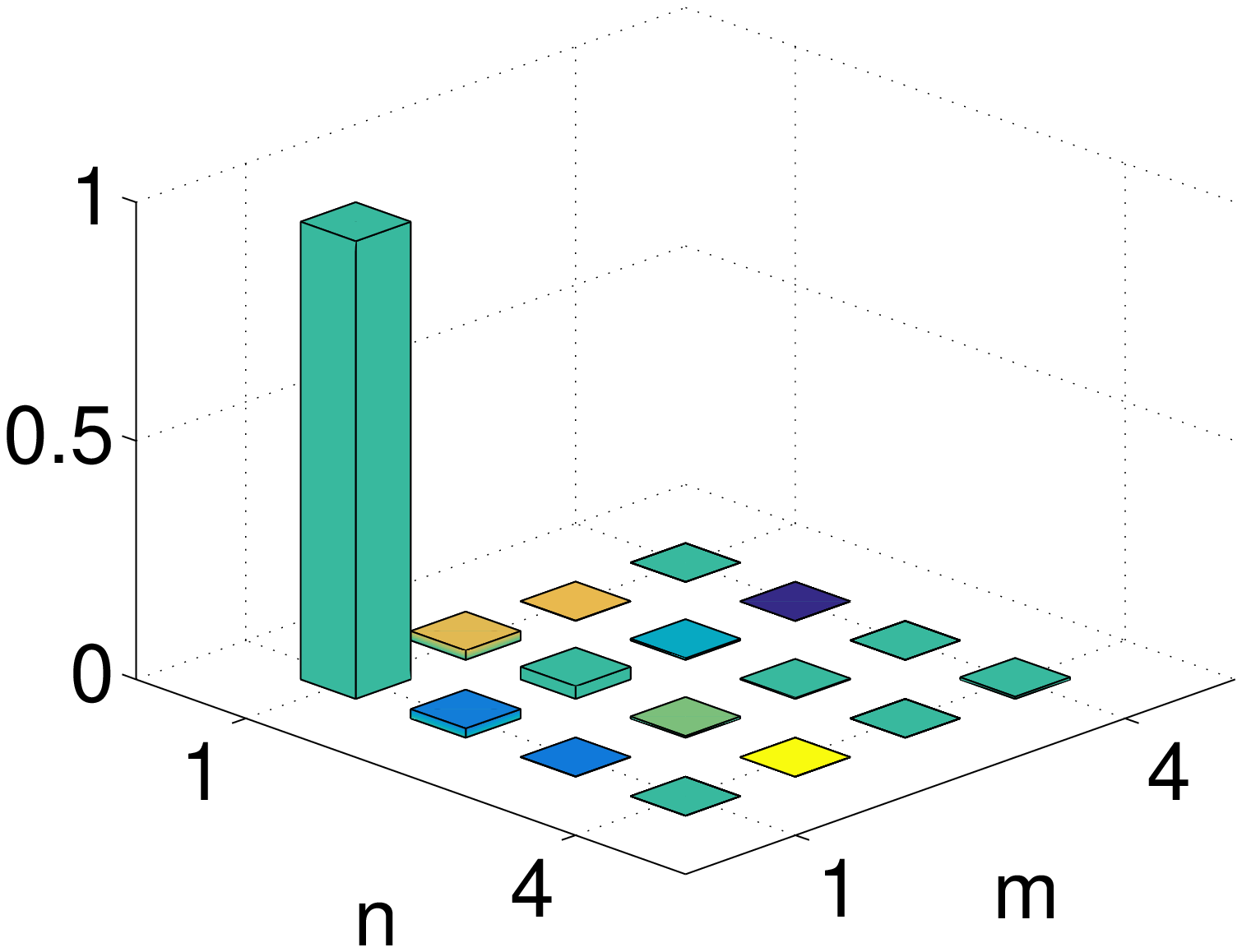}}
\subfloat[$\rho_{23}$]{
	\includegraphics[width=.15\textwidth]{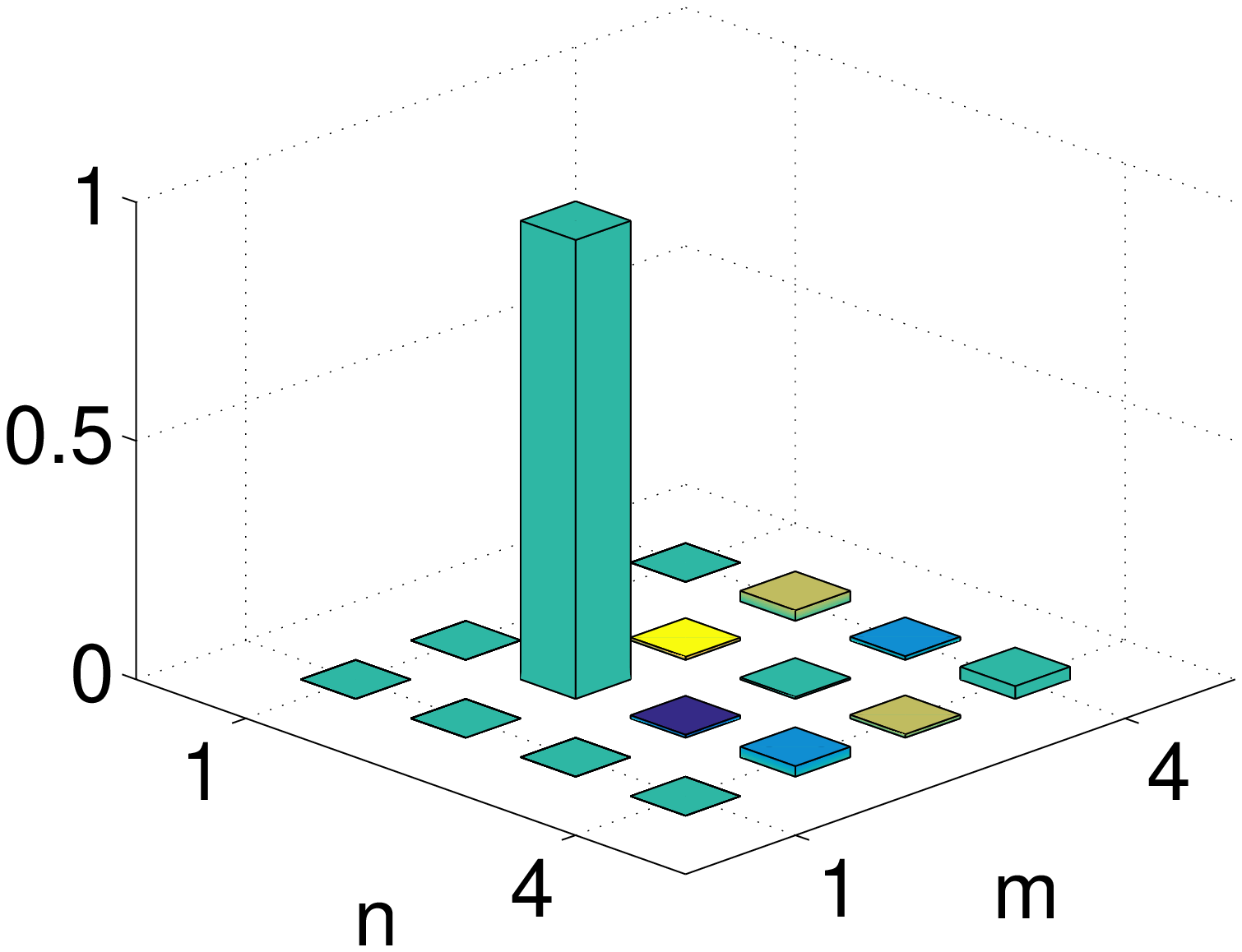}}
\caption{(color online) Calculated density matrices of the subsystems for the tripartite system.}
\end{figure}
  
After calculations, the strong subadditivity condition (Eq.\eqref{eq:strongsubadd}) reads: $0.2997 \le 0.3142 $; so the mutual information equals: $ I = S_{\rho_{12}} + S_{\rho_{23}} - S_{\rho} - S_{R_2} = 0.3142-0.2997 = 0.0147 $.

\section{\label{sec:concl}Conclusions}
We have checked that the experimentally measured density matix of a superconducting qudit \cite{shalibo2013direct} satisfies the new entropic inequalities for non-composite systems, given by equations (\ref{eq:subadd1}), (\ref{eq:subadd2}) and (\ref{eq:strongsubadd}). These inequalities can be further used to evaluate the accuracy of the experimental data.  Moreover, the value of mutual information, deduced from entropic inequalities, may charachterize correlations between different degrees of freedom in a noncomposite system. There also exist other inequalities for the von-Neumann and q-entropy, which will be checked in future publications.\\

\begin{acknowledgements}
We would like to thank E. Kiktenko and I. Khrapach for stimulating discussions.
\end{acknowledgements}

\bibliography{main}

\begin{thebibliography}{26}%
\makeatletter
\providecommand \@ifxundefined [1]{%
 \@ifx{#1\undefined}
}%
\providecommand \@ifnum [1]{%
 \ifnum #1\expandafter \@firstoftwo
 \else \expandafter \@secondoftwo
 \fi
}%
\providecommand \@ifx [1]{%
 \ifx #1\expandafter \@firstoftwo
 \else \expandafter \@secondoftwo
 \fi
}%
\providecommand \natexlab [1]{#1}%
\providecommand \enquote  [1]{``#1''}%
\providecommand \bibnamefont  [1]{#1}%
\providecommand \bibfnamefont [1]{#1}%
\providecommand \citenamefont [1]{#1}%
\providecommand \href@noop [0]{\@secondoftwo}%
\providecommand \href [0]{\begingroup \@sanitize@url \@href}%
\providecommand \@href[1]{\@@startlink{#1}\@@href}%
\providecommand \@@href[1]{\endgroup#1\@@endlink}%
\providecommand \@sanitize@url [0]{\catcode `\\12\catcode `\$12\catcode
  `\&12\catcode `\#12\catcode `\^12\catcode `\_12\catcode `\%12\relax}%
\providecommand \@@startlink[1]{}%
\providecommand \@@endlink[0]{}%
\providecommand \url  [0]{\begingroup\@sanitize@url \@url }%
\providecommand \@url [1]{\endgroup\@href {#1}{\urlprefix }}%
\providecommand \urlprefix  [0]{URL }%
\providecommand \Eprint [0]{\href }%
\providecommand \doibase [0]{http://dx.doi.org/}%
\providecommand \selectlanguage [0]{\@gobble}%
\providecommand \bibinfo  [0]{\@secondoftwo}%
\providecommand \bibfield  [0]{\@secondoftwo}%
\providecommand \translation [1]{[#1]}%
\providecommand \BibitemOpen [0]{}%
\providecommand \bibitemStop [0]{}%
\providecommand \bibitemNoStop [0]{.\EOS\space}%
\providecommand \EOS [0]{\spacefactor3000\relax}%
\providecommand \BibitemShut  [1]{\csname bibitem#1\endcsname}%
\let\auto@bib@innerbib\@empty
\bibitem [{\citenamefont {Dodonov}\ \emph {et~al.}(1989)\citenamefont
  {Dodonov}, \citenamefont {Man'ko},\ and\ \citenamefont
  {Man'ko}}]{dodonov1989correlated}%
  \BibitemOpen
  \bibfield  {author} {\bibinfo {author} {\bibfnamefont {V.}~\bibnamefont
  {Dodonov}}, \bibinfo {author} {\bibfnamefont {V.}~\bibnamefont {Man'ko}}, \
  and\ \bibinfo {author} {\bibfnamefont {O.}~\bibnamefont {Man'ko}},\
  }\href@noop {} {\bibfield  {journal} {\bibinfo  {journal} {Journal of Soviet
  Laser Research}\ }\textbf {\bibinfo {volume} {10}},\ \bibinfo {pages} {413}
  (\bibinfo {year} {1989})}\BibitemShut {NoStop}%
\bibitem [{\citenamefont {Fujii}\ \emph {et~al.}(2011)\citenamefont {Fujii},
  \citenamefont {Matsuo}, \citenamefont {Hatakenaka}, \citenamefont
  {Kurihara},\ and\ \citenamefont {Zeilinger}}]{fujii2011quantum}%
  \BibitemOpen
  \bibfield  {author} {\bibinfo {author} {\bibfnamefont {T.}~\bibnamefont
  {Fujii}}, \bibinfo {author} {\bibfnamefont {S.}~\bibnamefont {Matsuo}},
  \bibinfo {author} {\bibfnamefont {N.}~\bibnamefont {Hatakenaka}}, \bibinfo
  {author} {\bibfnamefont {S.}~\bibnamefont {Kurihara}}, \ and\ \bibinfo
  {author} {\bibfnamefont {A.}~\bibnamefont {Zeilinger}},\ }\href@noop {}
  {\bibfield  {journal} {\bibinfo  {journal} {Physical Review B}\ }\textbf
  {\bibinfo {volume} {84}},\ \bibinfo {pages} {174521} (\bibinfo {year}
  {2011})}\BibitemShut {NoStop}%
\bibitem [{\citenamefont {Clarke}\ and\ \citenamefont
  {Wilhelm}(2008)}]{clarke2008superconducting}%
  \BibitemOpen
  \bibfield  {author} {\bibinfo {author} {\bibfnamefont {J.}~\bibnamefont
  {Clarke}}\ and\ \bibinfo {author} {\bibfnamefont {F.~K.}\ \bibnamefont
  {Wilhelm}},\ }\href@noop {} {\bibfield  {journal} {\bibinfo  {journal}
  {Nature}\ }\textbf {\bibinfo {volume} {453}},\ \bibinfo {pages} {1031}
  (\bibinfo {year} {2008})}\BibitemShut {NoStop}%
\bibitem [{\citenamefont {Nakamura}\ \emph {et~al.}(1997)\citenamefont
  {Nakamura}, \citenamefont {Chen},\ and\ \citenamefont
  {Tsai}}]{nakamura1997spectroscopy}%
  \BibitemOpen
  \bibfield  {author} {\bibinfo {author} {\bibfnamefont {Y.}~\bibnamefont
  {Nakamura}}, \bibinfo {author} {\bibfnamefont {C.}~\bibnamefont {Chen}}, \
  and\ \bibinfo {author} {\bibfnamefont {J.}~\bibnamefont {Tsai}},\ }\href@noop
  {} {\bibfield  {journal} {\bibinfo  {journal} {Physical Review Letters}\
  }\textbf {\bibinfo {volume} {79}},\ \bibinfo {pages} {2328} (\bibinfo {year}
  {1997})}\BibitemShut {NoStop}%
\bibitem [{\citenamefont {Averkin}\ \emph {et~al.}(2014)\citenamefont
  {Averkin}, \citenamefont {Karpov}, \citenamefont {Shulga}, \citenamefont
  {Glushkov}, \citenamefont {Abramov}, \citenamefont {Huebner}, \citenamefont
  {Il'ichev},\ and\ \citenamefont {Ustinov}}]{averkin2014broadband}%
  \BibitemOpen
  \bibfield  {author} {\bibinfo {author} {\bibfnamefont {A.}~\bibnamefont
  {Averkin}}, \bibinfo {author} {\bibfnamefont {A.}~\bibnamefont {Karpov}},
  \bibinfo {author} {\bibfnamefont {K.}~\bibnamefont {Shulga}}, \bibinfo
  {author} {\bibfnamefont {E.}~\bibnamefont {Glushkov}}, \bibinfo {author}
  {\bibfnamefont {N.}~\bibnamefont {Abramov}}, \bibinfo {author} {\bibfnamefont
  {U.}~\bibnamefont {Huebner}}, \bibinfo {author} {\bibfnamefont
  {E.}~\bibnamefont {Il'ichev}}, \ and\ \bibinfo {author} {\bibfnamefont
  {A.}~\bibnamefont {Ustinov}},\ }\href@noop {} {\bibfield  {journal} {\bibinfo
   {journal} {Review of Scientific Instruments}\ }\textbf {\bibinfo {volume}
  {85}},\ \bibinfo {pages} {104702} (\bibinfo {year} {2014})}\BibitemShut
  {NoStop}%
\bibitem [{\citenamefont {Chiorescu}\ \emph {et~al.}(2003)\citenamefont
  {Chiorescu}, \citenamefont {Nakamura}, \citenamefont {Harmans},\ and\
  \citenamefont {Mooij}}]{chiorescu2003coherent}%
  \BibitemOpen
  \bibfield  {author} {\bibinfo {author} {\bibfnamefont {I.}~\bibnamefont
  {Chiorescu}}, \bibinfo {author} {\bibfnamefont {Y.}~\bibnamefont {Nakamura}},
  \bibinfo {author} {\bibfnamefont {C.~M.}\ \bibnamefont {Harmans}}, \ and\
  \bibinfo {author} {\bibfnamefont {J.}~\bibnamefont {Mooij}},\ }\href@noop {}
  {\bibfield  {journal} {\bibinfo  {journal} {Science}\ }\textbf {\bibinfo
  {volume} {299}},\ \bibinfo {pages} {1869} (\bibinfo {year}
  {2003})}\BibitemShut {NoStop}%
\bibitem [{\citenamefont {Steffen}\ \emph {et~al.}(2006)\citenamefont
  {Steffen}, \citenamefont {Ansmann}, \citenamefont {McDermott}, \citenamefont
  {Katz}, \citenamefont {Bialczak}, \citenamefont {Lucero}, \citenamefont
  {Neeley}, \citenamefont {Weig}, \citenamefont {Cleland},\ and\ \citenamefont
  {Martinis}}]{steffen2006state}%
  \BibitemOpen
  \bibfield  {author} {\bibinfo {author} {\bibfnamefont {M.}~\bibnamefont
  {Steffen}}, \bibinfo {author} {\bibfnamefont {M.}~\bibnamefont {Ansmann}},
  \bibinfo {author} {\bibfnamefont {R.}~\bibnamefont {McDermott}}, \bibinfo
  {author} {\bibfnamefont {N.}~\bibnamefont {Katz}}, \bibinfo {author}
  {\bibfnamefont {R.~C.}\ \bibnamefont {Bialczak}}, \bibinfo {author}
  {\bibfnamefont {E.}~\bibnamefont {Lucero}}, \bibinfo {author} {\bibfnamefont
  {M.}~\bibnamefont {Neeley}}, \bibinfo {author} {\bibfnamefont {E.~M.}\
  \bibnamefont {Weig}}, \bibinfo {author} {\bibfnamefont {A.~N.}\ \bibnamefont
  {Cleland}}, \ and\ \bibinfo {author} {\bibfnamefont {J.~M.}\ \bibnamefont
  {Martinis}},\ }\href@noop {} {\bibfield  {journal} {\bibinfo  {journal}
  {Physical Review Letters}\ }\textbf {\bibinfo {volume} {97}},\ \bibinfo
  {pages} {050502} (\bibinfo {year} {2006})}\BibitemShut {NoStop}%
\bibitem [{\citenamefont {Shalibo}\ \emph {et~al.}(2013)\citenamefont
  {Shalibo}, \citenamefont {Resh}, \citenamefont {Fogel}, \citenamefont {Shwa},
  \citenamefont {Bialczak}, \citenamefont {Martinis},\ and\ \citenamefont
  {Katz}}]{shalibo2013direct}%
  \BibitemOpen
  \bibfield  {author} {\bibinfo {author} {\bibfnamefont {Y.}~\bibnamefont
  {Shalibo}}, \bibinfo {author} {\bibfnamefont {R.}~\bibnamefont {Resh}},
  \bibinfo {author} {\bibfnamefont {O.}~\bibnamefont {Fogel}}, \bibinfo
  {author} {\bibfnamefont {D.}~\bibnamefont {Shwa}}, \bibinfo {author}
  {\bibfnamefont {R.}~\bibnamefont {Bialczak}}, \bibinfo {author}
  {\bibfnamefont {J.~M.}\ \bibnamefont {Martinis}}, \ and\ \bibinfo {author}
  {\bibfnamefont {N.}~\bibnamefont {Katz}},\ }\href@noop {} {\bibfield
  {journal} {\bibinfo  {journal} {Physical Review Letters}\ }\textbf {\bibinfo
  {volume} {110}},\ \bibinfo {pages} {100404} (\bibinfo {year}
  {2013})}\BibitemShut {NoStop}%
\bibitem [{\citenamefont {Shannon}(1948)}]{shannon1948note}%
  \BibitemOpen
  \bibfield  {author} {\bibinfo {author} {\bibfnamefont {C.~E.}\ \bibnamefont
  {Shannon}},\ }\href@noop {} {\bibfield  {journal} {\bibinfo  {journal} {Bell
  System Tech. J}\ }\textbf {\bibinfo {volume} {27}},\ \bibinfo {pages} {379}
  (\bibinfo {year} {1948})}\BibitemShut {NoStop}%
\bibitem [{\citenamefont {Holevo}(2012)}]{holevo2012quantum}%
  \BibitemOpen
  \bibfield  {author} {\bibinfo {author} {\bibfnamefont {A.~S.}\ \bibnamefont
  {Holevo}},\ }\href@noop {} {\emph {\bibinfo {title} {Quantum systems,
  channels, information: a mathematical introduction}}},\ Vol.~\bibinfo
  {volume} {16}\ (\bibinfo  {publisher} {Walter de Gruyter},\ \bibinfo {year}
  {2012})\BibitemShut {NoStop}%
\bibitem [{\citenamefont {Fedorov}\ \emph {et~al.}(2014)\citenamefont
  {Fedorov}, \citenamefont {Kiktenko}, \citenamefont {Man'ko},\ and\
  \citenamefont {Man'ko}}]{fedorov2014entropic}%
  \BibitemOpen
  \bibfield  {author} {\bibinfo {author} {\bibfnamefont {A.}~\bibnamefont
  {Fedorov}}, \bibinfo {author} {\bibfnamefont {E.}~\bibnamefont {Kiktenko}},
  \bibinfo {author} {\bibfnamefont {O.}~\bibnamefont {Man'ko}}, \ and\ \bibinfo
  {author} {\bibfnamefont {V.}~\bibnamefont {Man'ko}},\ }\href@noop {}
  {\bibfield  {journal} {\bibinfo  {journal} {arXiv preprint arXiv:1411.0157}\
  } (\bibinfo {year} {2014})}\BibitemShut {NoStop}%
\bibitem [{\citenamefont {Kiktenko}\ \emph
  {et~al.}(2015{\natexlab{a}})\citenamefont {Kiktenko}, \citenamefont
  {Fedorov}, \citenamefont {Man'ko},\ and\ \citenamefont
  {Man'ko}}]{kiktenko2015multilevel}%
  \BibitemOpen
  \bibfield  {author} {\bibinfo {author} {\bibfnamefont {E.}~\bibnamefont
  {Kiktenko}}, \bibinfo {author} {\bibfnamefont {A.}~\bibnamefont {Fedorov}},
  \bibinfo {author} {\bibfnamefont {O.}~\bibnamefont {Man'ko}}, \ and\ \bibinfo
  {author} {\bibfnamefont {V.}~\bibnamefont {Man'ko}},\ }\href@noop {}
  {\bibfield  {journal} {\bibinfo  {journal} {Physical Review A}\ }\textbf
  {\bibinfo {volume} {91}},\ \bibinfo {pages} {042312} (\bibinfo {year}
  {2015}{\natexlab{a}})}\BibitemShut {NoStop}%
\bibitem [{\citenamefont {Kiktenko}\ \emph
  {et~al.}(2015{\natexlab{b}})\citenamefont {Kiktenko}, \citenamefont
  {Fedorov}, \citenamefont {Strakhov},\ and\ \citenamefont
  {Man'ko}}]{kiktenko2015single}%
  \BibitemOpen
  \bibfield  {author} {\bibinfo {author} {\bibfnamefont {E.}~\bibnamefont
  {Kiktenko}}, \bibinfo {author} {\bibfnamefont {A.}~\bibnamefont {Fedorov}},
  \bibinfo {author} {\bibfnamefont {A.}~\bibnamefont {Strakhov}}, \ and\
  \bibinfo {author} {\bibfnamefont {V.}~\bibnamefont {Man'ko}},\ }\href@noop {}
  {\bibfield  {journal} {\bibinfo  {journal} {Physics Letters A}\ }\textbf
  {\bibinfo {volume} {379}},\ \bibinfo {pages} {1409} (\bibinfo {year}
  {2015}{\natexlab{b}})}\BibitemShut {NoStop}%
\bibitem [{\citenamefont {Man'ko}\ and\ \citenamefont
  {Man'ko}(2014{\natexlab{a}})}]{man2014quantum}%
  \BibitemOpen
  \bibfield  {author} {\bibinfo {author} {\bibfnamefont {M.~A.}\ \bibnamefont
  {Man'ko}}\ and\ \bibinfo {author} {\bibfnamefont {V.~I.}\ \bibnamefont
  {Man'ko}},\ }\href@noop {} {\bibfield  {journal} {\bibinfo  {journal}
  {Physica Scripta}\ }\textbf {\bibinfo {volume} {2014}},\ \bibinfo {pages}
  {014030} (\bibinfo {year} {2014}{\natexlab{a}})}\BibitemShut {NoStop}%
\bibitem [{\citenamefont {Man’ko}\ and\ \citenamefont
  {Markovich}(2014)}]{man2014separability}%
  \BibitemOpen
  \bibfield  {author} {\bibinfo {author} {\bibfnamefont {V.~I.}\ \bibnamefont
  {Man’ko}}\ and\ \bibinfo {author} {\bibfnamefont {L.}~\bibnamefont
  {Markovich}},\ }\href@noop {} {\bibfield  {journal} {\bibinfo  {journal}
  {Journal of Russian Laser Research}\ }\textbf {\bibinfo {volume} {35}},\
  \bibinfo {pages} {518} (\bibinfo {year} {2014})}\BibitemShut {NoStop}%
\bibitem [{\citenamefont {Chernega}\ and\ \citenamefont
  {Man’ko}(2014)}]{chernega2014tomographic}%
  \BibitemOpen
  \bibfield  {author} {\bibinfo {author} {\bibfnamefont {V.~N.}\ \bibnamefont
  {Chernega}}\ and\ \bibinfo {author} {\bibfnamefont {O.~V.}\ \bibnamefont
  {Man’ko}},\ }\href@noop {} {\bibfield  {journal} {\bibinfo  {journal}
  {Journal of Russian Laser Research}\ }\textbf {\bibinfo {volume} {35}},\
  \bibinfo {pages} {27} (\bibinfo {year} {2014})}\BibitemShut {NoStop}%
\bibitem [{\citenamefont {Chernega}\ \emph {et~al.}(2014)\citenamefont
  {Chernega}, \citenamefont {Man'ko},\ and\ \citenamefont
  {Man'ko}}]{chernega2014deformed}%
  \BibitemOpen
  \bibfield  {author} {\bibinfo {author} {\bibfnamefont {V.~N.}\ \bibnamefont
  {Chernega}}, \bibinfo {author} {\bibfnamefont {O.~V.}\ \bibnamefont
  {Man'ko}}, \ and\ \bibinfo {author} {\bibfnamefont {V.~I.}\ \bibnamefont
  {Man'ko}},\ }\href@noop {} {\bibfield  {journal} {\bibinfo  {journal} {arXiv
  preprint arXiv:1412.6771}\ } (\bibinfo {year} {2014})}\BibitemShut {NoStop}%
\bibitem [{\citenamefont {Man'ko}\ and\ \citenamefont
  {Man'ko}(2014{\natexlab{b}})}]{man2014entanglement}%
  \BibitemOpen
  \bibfield  {author} {\bibinfo {author} {\bibfnamefont {M.~A.}\ \bibnamefont
  {Man'ko}}\ and\ \bibinfo {author} {\bibfnamefont {V.~I.}\ \bibnamefont
  {Man'ko}},\ }\href@noop {} {\bibfield  {journal} {\bibinfo  {journal}
  {International Journal of Quantum Information}\ } (\bibinfo {year}
  {2014}{\natexlab{b}})}\BibitemShut {NoStop}%
\bibitem [{\citenamefont {Chernega}\ and\ \citenamefont
  {Man'ko}(2015)}]{chernega2015no}%
  \BibitemOpen
  \bibfield  {author} {\bibinfo {author} {\bibfnamefont {V.}~\bibnamefont
  {Chernega}}\ and\ \bibinfo {author} {\bibfnamefont {O.}~\bibnamefont
  {Man'ko}},\ }\href@noop {} {\bibfield  {journal} {\bibinfo  {journal} {arXiv
  preprint arXiv:1504.03858, Physica Scripta accepted}\ } (\bibinfo {year}
  {2015})}\BibitemShut {NoStop}%
\bibitem [{\citenamefont {Shalibo}(2012)}]{shalibo2012control}%
  \BibitemOpen
  \bibfield  {author} {\bibinfo {author} {\bibfnamefont {Y.~P.}\ \bibnamefont
  {Shalibo}},\ }\emph {\bibinfo {title} {Control and Measurement of Multi-level
  States in the Josephson Phase Circuit}},\ \href@noop {} {Ph.D. thesis},\
  \bibinfo  {school} {Hebrew University of Jerusalem} (\bibinfo {year}
  {2012})\BibitemShut {NoStop}%
\bibitem [{\citenamefont {Katz}\ and\ \citenamefont
  {Shalibo}()}]{katz2015private}%
  \BibitemOpen
  \bibfield  {author} {\bibinfo {author} {\bibfnamefont {N.}~\bibnamefont
  {Katz}}\ and\ \bibinfo {author} {\bibfnamefont {Y.}~\bibnamefont {Shalibo}},\
  }\href@noop {} {}\bibinfo {howpublished} {personal communication}\BibitemShut
  {NoStop}%
\bibitem [{\citenamefont {Josephson}(1962)}]{josephson1962possible}%
  \BibitemOpen
  \bibfield  {author} {\bibinfo {author} {\bibfnamefont {B.~D.}\ \bibnamefont
  {Josephson}},\ }\href@noop {} {\bibfield  {journal} {\bibinfo  {journal}
  {Physics Letters}\ }\textbf {\bibinfo {volume} {1}},\ \bibinfo {pages} {251}
  (\bibinfo {year} {1962})}\BibitemShut {NoStop}%
\bibitem [{\citenamefont {Martinis}\ \emph {et~al.}(2002)\citenamefont
  {Martinis}, \citenamefont {Nam}, \citenamefont {Aumentado},\ and\
  \citenamefont {Urbina}}]{martinis2002rabi}%
  \BibitemOpen
  \bibfield  {author} {\bibinfo {author} {\bibfnamefont {J.~M.}\ \bibnamefont
  {Martinis}}, \bibinfo {author} {\bibfnamefont {S.}~\bibnamefont {Nam}},
  \bibinfo {author} {\bibfnamefont {J.}~\bibnamefont {Aumentado}}, \ and\
  \bibinfo {author} {\bibfnamefont {C.}~\bibnamefont {Urbina}},\ }\href@noop {}
  {\bibfield  {journal} {\bibinfo  {journal} {Physical Review Letters}\
  }\textbf {\bibinfo {volume} {89}},\ \bibinfo {pages} {117901} (\bibinfo
  {year} {2002})}\BibitemShut {NoStop}%
\bibitem [{\citenamefont {Neeley}\ \emph {et~al.}(2009)\citenamefont {Neeley},
  \citenamefont {Ansmann}, \citenamefont {Bialczak}, \citenamefont {Hofheinz},
  \citenamefont {Lucero}, \citenamefont {O'Connell}, \citenamefont {Sank},
  \citenamefont {Wang}, \citenamefont {Wenner}, \citenamefont {Cleland} \emph
  {et~al.}}]{neeley2009emulation}%
  \BibitemOpen
  \bibfield  {author} {\bibinfo {author} {\bibfnamefont {M.}~\bibnamefont
  {Neeley}}, \bibinfo {author} {\bibfnamefont {M.}~\bibnamefont {Ansmann}},
  \bibinfo {author} {\bibfnamefont {R.~C.}\ \bibnamefont {Bialczak}}, \bibinfo
  {author} {\bibfnamefont {M.}~\bibnamefont {Hofheinz}}, \bibinfo {author}
  {\bibfnamefont {E.}~\bibnamefont {Lucero}}, \bibinfo {author} {\bibfnamefont
  {A.~D.}\ \bibnamefont {O'Connell}}, \bibinfo {author} {\bibfnamefont
  {D.}~\bibnamefont {Sank}}, \bibinfo {author} {\bibfnamefont {H.}~\bibnamefont
  {Wang}}, \bibinfo {author} {\bibfnamefont {J.}~\bibnamefont {Wenner}},
  \bibinfo {author} {\bibfnamefont {A.~N.}\ \bibnamefont {Cleland}},  \emph
  {et~al.},\ }\href@noop {} {\bibfield  {journal} {\bibinfo  {journal}
  {Science}\ }\textbf {\bibinfo {volume} {325}},\ \bibinfo {pages} {722}
  (\bibinfo {year} {2009})}\BibitemShut {NoStop}%
\bibitem [{\citenamefont {Neumann}(1955)}]{john1955mathematical}%
  \BibitemOpen
  \bibfield  {author} {\bibinfo {author} {\bibfnamefont {J.~V.}\ \bibnamefont
  {Neumann}},\ }\href@noop {} {\emph {\bibinfo {title} {Mathematical
  foundations of quantum mechanics}}},\ \bibinfo {number} {2}\ (\bibinfo
  {publisher} {Princeton university press},\ \bibinfo {year}
  {1955})\BibitemShut {NoStop}%
\bibitem [{\citenamefont {Lieb}\ and\ \citenamefont
  {Ruskai}(2002)}]{lieb2002proof}%
  \BibitemOpen
  \bibfield  {author} {\bibinfo {author} {\bibfnamefont {E.~H.}\ \bibnamefont
  {Lieb}}\ and\ \bibinfo {author} {\bibfnamefont {M.~B.}\ \bibnamefont
  {Ruskai}},\ }in\ \href@noop {} {\emph {\bibinfo {booktitle} {Inequalities}}}\
  (\bibinfo  {publisher} {Springer},\ \bibinfo {year} {2002})\ pp.\ \bibinfo
  {pages} {63--66}\BibitemShut {NoStop}%
\end{thebibliography}%

\end{document}